\documentclass[prd, showpacs, showkeys, preprintnumbers]{revtex4}
\usepackage{graphicx}
\usepackage{dcolumn}
\usepackage{bm}

\newcommand{\be}{\begin{equation}}
\newcommand{\ee}{\end{equation}}
\newcommand{\bea}{\begin{eqnarray}}
\newcommand{\eea}{\end{eqnarray}}
\newcommand{\ba}{\begin{array}}
\newcommand{\ea}{\end{array}}
\newcommand{\lsim}
{{\;\raise0.3ex\hbox{$<$\kern-0.75em\raise-1.1ex\hbox{$\sim$}}\;}}
\newcommand{\gsim}
{{\;\raise0.3ex\hbox{$>$\kern-0.75em\raise-1.1ex\hbox{$\sim$}}\;}}
\def\msQ{m^2_{\tilde Q_L}}
\def\msuL{m^2_{\tilde u_L}}
\def\msdL{m^2_{\tilde d_L}}
\def\msuR{m^2_{\tilde u_R}}
\def\msdR{m^2_{\tilde d_R}}
\def\mslL{m^2_{\tilde L_L}}
\def\mseL{m^2_{\tilde e_L}}
\def\msnL{m^2_{\tilde{\nu}_L}}
\def\mseR{m^2_{\tilde e_R}}
\def\mhU{m^2_{H_u}}
\def\mhD{m^2_{H_d}}

\def\st{\sin^2 \theta_W}

\begin{document}

\bigskip

\preprint{IISc-CHEP-4/04}
\preprint{\tt{hep-ph/0412125}}

\bigskip

\bigskip


\title{\bf Probing  $SO(10)$ symmetry breaking patterns through
sfermion  mass relations}

\vspace{1.5in}

\bigskip

\author{B. Ananthanarayan}

\vspace{1in}

\affiliation{
Centre for High Energy Physics, Indian Institute of 
Science, Bangalore 560 012, India} 

\vspace{2mm}

\author{P. N. Pandita}

\affiliation{Department of Physics,
North-Eastern Hill University,  Shillong 793 022, India}

\begin{abstract}


We consider supersymmetric $SO(10)$ grand unification where the 
unified gauge group can break to the Standard Model gauge group 
through different chains.  The breaking of  $SO(10)$ necessarily involves 
the reduction of the rank, and consequent generation of non-universal 
supersymmetry breaking scalar mass terms.
We derive squark and slepton mass relations, taking into account
these non-universal contributions to the sfermion masses, 
which can help distinguish between the different chains through which 
the $SO(10)$ gauge group breaks to the Standard Model gauge group. 
We then study some implications of these non-universal
supersymmetry breaking scalar masses for the low energy phenomenology. 

\end{abstract}

\vspace{1cm}


\maketitle

\section{\label{sec:intro} Introduction}
Despite its stupendous success, the gauge group $SU(3)\times
SU(2)\times U(1)$ remains a completely unexplained feature of
the Standard Model~(SM) of electroweak and strong interactions. 
The idea of grand unification~\cite{GG1} 
is, therefore,  one of the most compelling theoretical
ideas that goes beyond the Standard Model.
In grand unified theories~(GUT's), the SM gauge group
can be elegantly unified into a simple group. Moreover, the fermion
content of the SM model can be accomodated in  irreducible
representations of the unified gauge group. Also, one can understand
the smallness of neutrino masses via the  seesaw mechanism~\cite{Gell-Mann:vs}
in some of the grand unified  models  like $SO(10)$ ~\cite{SO10} .
The renormalization group flow of the gauge couplings leads to
their unification at a very large scale ~\cite{GQW}.

This picture of physics beyond the SM
leads to the well-known hierarchy problem due to the
widely separated scales, the weak scale $\sim M_Z$,
and the large unification scale characterizing by the gauge coupling 
unification.  It has been argued that in supersymmetric~\cite{wess}
extensions of the standard model~\cite{Nilles}, the hierarchy between the 
two scales can be made technically natural.
This leads us to the idea  of supersymmetric grand
unification.  In supersymmetric GUTS, supersymmetry raises the GUT
prediction for $\sin^2\theta_W$~\cite{DRW}, which becomes very close to the
current measurement.
One of the most important predictions of grand unification is 
that, because of the presence of baryon number violating
interactions, the proton must decay. 
Since supersymmetry  raises the GUT scale,
the proton lifetime can be long enough to be consistent with
experiment~\cite{nathraby}.  

Presently the hope is that most of the supersymmetric particle 
spectrum would be observed at the Large Hadron Collider~(LHC).
One could then have detailed information on the properties, 
especially the masses,  production mechanisms and
decays of the sparticles either at LHC or a future Linear Collider~(LC). 
The question can then be posed as to what one can
infer from this information about grand unification, and
in particular, whether the masses of the sparticles, and their 
interrelationships, can provide us with a clue as to the
nature of the grand unified gauge group and its spontaneous 
breaking to the SM gauge group. Indeed one may go as far as to ask 
whether the pattern of sparticle  masses can rule out simple 
grand unification.

In this work we try to address this question in detail. We recall
that the simplest grand unified theory into which the SM can be embedded
is the $SU(5)$ grand unified theory~\cite{GG1}.  
The rank of $SU(5)$ is the same  as that of the 
SM gauge group.  On the other hand, SM can also be embedded 
into a larger gauge group like $SO(10)$.
However, since the rank of  $SO(10)$ is higher than the 
SM gauge group, the breaking of $SO(10)$ to the SM gauge group involves
the reduction of the rank by  one unit. Thus,  in the case of 
$SO(10)$ unification there will be  
$D-$term contributions to the soft supersymmetry breaking scalar masses.
In general $D-$term contributions
to the SUSY breaking soft scalar masses arise whenever a gauge symmetry is
spontaneously broken with a reduction of rank~\cite{Drees}.
These $D-$term contributions have important phenomenological
consequences at low energies
as they allow one to reach certain regions of parameter space which are
not otherwise accessible with  universal boundary
conditions~\cite{KM1,Autoetal}.
Non-universality, and in particular
the $D-$term contributions,  may have a dramatic
impact on certain sum rules~\cite{MR1}
satisfied by the squark and slepton masses. Such effects are likely
to help distinguish between different scenarios for breaking of grand unified
symmetry at high energies~\cite{KMYPLB,KMYPRD,CH1}.

In a recent work~\cite{bapnp} we addressed the question of 
D-term non-universality in the context of $SO(10)$ unified gauge 
group when it breaks to the SM  gauge group
via one of its maximal subgroups $SU(5)\times U(1)$.  
In the present paper we systematically consider the $D-$ term non-universality
that is generated in  $SO(10)$ unification when it breaks to the SM gauge group
via any of its maximal subgroups. Since no such contributions
are generated in $SU(5)$ unification, $SO(10)$ is one of the two
($E_6$ being the other) supersymmetric grand unified theories in four
dimensions where such contributions can arise. In 
Section~\ref{section:so10break}, we discuss in detail the embedding of
the Standard Model in $SO(10)$ grand unified gauge group, and study 
the different chains through which it can break to the SM  
gauge group. Here we  discuss  why the embedding of SM into $SO(10)$
can be done in more than one way.
In Section~\ref{section:Rgsumrules} we consider the renormalization group
equations and their solutions for  $SO(10)$ breaking into the SM for 
different chains of breaking. Using these solutions, we derive characteristic
relations between the sfermion masses which hold for different
patterns of  $SO(10)$ breaking into the SM gauge group. 
In Section~\ref{section:Numerical} we carry out a numerical analysis of the
renormalization group evolution for the  $SO(10)$ breaking into the 
SM gauge  group, and the implications
of this evolution for the low energy phenomenology.
In this Section we also  address the question of  naturalness of the large
values of $\tan \beta$ in the context of grand unified 
$SO(10)$ models with  non-universal $D$-term
contributions as well as certain other kinds of
non-universality. We conclude the paper with a summary and some remarks.

\section{\label{section:so10break} Embedding of SM in $SO(10)$}
As pointed out in the Introduction, the SM can be embedded into a larger
gauge group, where an entire SM generation can be fitted into
a single irreducible representation of the underlying gauge group.
Indeed, there is chain of group embeddings~\cite{Witten:2002ei}
\be
{SU(5)\subset SO(10)\subset E_6\subset E_7\subset E_8.}
\label{embed1}
\ee
However, in four-dimensional grand unified theories the gauge groups
$E_7$ and $E_8$ do not support a chiral structure
of the weak interactions, and hence cannot be
used as grand unified gauge groups. This leaves out only
the three groups,
$SU(5)$, $SO(10)$, and  $E_6$ as possible grand unified gauge
groups in four dimensions.
The gauge group  $SO(10)$ appears at present to be
the  most attractive because
it contains an entire SM generation  in the fundamental representation.
Furthermore,  one has as bonus the right handed neutrino, necessary
to generate neutrino masses, sitting in the same fundamental
representation.  Also,  a complex
${\bf 10}$ dimensional representation of $SO(10)$ can be employed
to accomodate the two Higgs doublets of the low energy minimal
supersymmetric standard model.
However, since the rank of $SO(10)$ is one unit higher than
the SM gauge group, it leads to $D$-term contributions to the 
soft scalar masses at the scale of symmetry breaking. These  $D$-term
contributions will depend on the manner  in which the  $SO(10)$
gauge group is broken to the SM gauge group. In order to 
study the implications of these  
$D$-term contributions for the phenomenology, we shall, in the following,
discuss the breaking of  $SO(10)$ in detail.

We start by recalling  that when $SO(10)$ breaks via its maximal
subgroup $SU(5)\times U(1)_Z$, with 
$SU(5) \supset SU(3)_C\times SU(2)_L \times U(1)_X$, there are
two possibilities for the hypercharge generator of the 
SM gauge group.
In the ``conventional'' embedding via $SU(5)$, 
the hypercharge generator $Y$ of the SM is identified with the generator 
$X$ of $U(1)_X$. On the other hand, in the ``flipped'' embedding
the hypercharge generator is identified with a linear combination of
the generators $X$ and $Z$. 


Apart from the ``natural'' subgroup 
$SU(5)\times U(1)$, the group $SO(10)$ also has ``natural'' 
subgroup $SO(6)\times SO(4).$
Since $SO(6)$ is isomorphic to $SU(4)$, and $SO(4)$  is isomorphic
to $SU(2) \times SU(2)$, $SO(10)$ contains~\cite{Pati:1974yy} the group
$SU(4)\times SU(2) \times SU(2)$. We shall focuss on the
signatures of the  $SO(10)$ breaking via its two natural subgroups, 
and try to find distinguishing features of the sparticle spectrum 
in the two cases.

In  $SO(10)$ grand unification, all the matter particles of one family
of the Standard Model~(SM) together with a right handed neutrino belong
to the spinor representation ${\bf 16}$.  Each such spinor representation
${\bf 16}$ and each ${\bf 10}$-dimensional representation can be decomposed 
under the maximal subgroup
$SO(10) \supset SU(5) \times U(1)_Z$ as
\bea
{\bf 16} &=& {\bf 5^*}_{3} + {\bf 10}_{-1} + {\bf 1}_{-5},\label{decomp16} \\
{\bf 10} &=& {\bf 5}_2 +{5^*}_{-2}.\label{decomp10}
\eea
Furthermore under $SU(5) \supset SU(3)_C \times SU(2)_L \times U(1)_X,$ 
we have the decomposition
\bea
{\bf 5}& = & ({\bf 3},{\bf 1})_{-2} + ({\bf 1},{\bf 2})_{3} , \label{decomp5} \\
{\bf 5^*}& = & ({\bf 3^*},{\bf 1})_2 + ({\bf 1},{\bf 2})_{-3} ,
\label{decomp5*} \\
{\bf 10} & = & ({\bf 3},{\bf 2})_1 + ({\bf 3^*},{\bf 1})_{-4}
+({\bf 1},{\bf 1})_6 , \label{decomp10p} \\
{\bf 1}  & = & ({\bf 1},{\bf 1})_0.\label{decomp1}
\eea
We note that $U(1)_X$, which is the subgroup of $SU(5)$, is not identical
with the $U(1)_Y$  of the SM at this stage. Note also that each 
${\bf16}$ includes two pairs of 
$({\bf 3^*},{\bf 1})$ and  $({\bf 1},{\bf 1}).$ 

In order to identify the hypercharge group, we consider the
decomposition $SO(10) \supset SU(5)\times U(1)_Z \supset
SU(3)_C \times SU(2)_L \times U(1)_X \times U(1)_Z$.
Therefore, the hypercharge $U(1)_Y$ must be a linear combination of
$U(1)_X$ and $U(1)_Z$, i.e. $U(1)_Y \subset U(1)_X \times U(1)_Z$.
Thus, there are two ways to define the  hypercharge generator of the
SM:
\bea
Y &=& X,  \label{gg1} \\
Y &=& -\frac{1}{5}(X+6Z) \label{flipped1},
\eea
upto an overall normalization factor.
The first case corresponds to the the Georgi-Glashow model~\cite{GG1},
whereas the second identification of hypercharge corresponds to the flipped
case~\cite{DRGG,SMB}.

In the first case the $U(1)$ generator of $SO(10)$ that is orthogonal to
$Y$ and the diagonal generators of $SU(3)_C$  and $SU(2)_L$ is
\bea
Y^{\bot} &=& -Z,
\eea
whereas in the flipped case we have for the orthogonal generator\cite{bapnp}
\bea
Y^{\bot} &=& {-4X+Z\over 5},
\eea
After a suitable identification of the fields lying in the relevant
representations of $SO(10)$, the effect of $SO(10)$ breaking at the unification
scale leads to $D$-term non-universality, which is computed in terms of
the eigenvalues of the operator $Y^\perp$ on the fields.
We will discuss this in the next section.

We now come to the case of $SO(10)$ breaking via the  Pati-Salam subgroup.
The breaking pattern to the SM gauge group is
\newcommand{\verylongrightarrow}{
\relbar\joinrel\relbar\joinrel\relbar\joinrel\rightarrow}
\newcommand{\breaksto}[1]{\mathop{\verylongrightarrow}\limits^{#1}}
\begin{equation}
SO(10) \breaksto{M_{U}} SU(4)_{PS} \times SU(2)_L \times SU(2)_R
        \breaksto{M_{PS}} SU(3)_C \times SU(2)_L \times U(1)_Y.
\end{equation}
The decomposition of {\bf 16} and ${\bf 10}$ of $SO(10)$ under the 
maximal subgroup $SO(10) \supset SU(4)_{PS} \times SU(2)_L \times SU(2)_R$ 
is given by
\begin{eqnarray}
{\bf 16} &=& ({\bf 4},{\bf 2},{\bf 1}) + ({\bf 4^*},{\bf 1},{\bf 2}),
\label{16su4}\\
{\bf 10} &=& ({\bf 6},{\bf 1},{\bf 1}) + ({\bf 1},{\bf 2},{\bf 2}).
\label{10su4}
\end{eqnarray}
Furthermore, under $SU(4)_{PS} \times SU(2)_L \times SU(2)_R \supset
SU(3)_C \times SU(2)_L \times SU(2)_R \times U(1)_V$ we have 
the decomposition
\begin{eqnarray}
({\bf 4},{\bf 2},{\bf 1}) & = & ({\bf 3},{\bf 2},{\bf 1})_{1/3}
                                +({\bf 1},{\bf 2},{\bf 1})_{-1} ,
                                 \label{decom421}\\
({\bf 4^*},{\bf 1},{\bf 2}) & = & ({\bf 3^*},{\bf 1},{\bf 2})_{-1/3}
                                +({\bf 1},{\bf 1},{\bf 2})_{1},
\label{decom412}\\
({\bf 6},{\bf 1},{\bf 1}) & = & ({\bf 3},{\bf 1},{\bf 1})_{-2/3}
                                +({\bf 3^*},{\bf 1},{\bf 1})_{2/3} ,
                                 \label{decom611} \\
({\bf 1},{\bf 2},{\bf 2}) & = & ({\bf 1},{\bf 2},{\bf 2})_0.
\label{decom122}
\end{eqnarray}
The decomposition~(\ref{decom412}) shows that 
$({\bf 3^*},{\bf 1})$ and $({\bf 1},{\bf 1})$ are $SU(2)_R$ doublets.
It is easily seen that the generators $U(1)_Z$ and $U(1)_X$,  and the
generator $U(1)_V$ are related through
\begin{eqnarray}
Z & = & -4I_{3R} -3V , \label{Znumber} \\
6X & = & -I_{3R} + \frac{1}{2}V \label{Xnumber}.
\end{eqnarray}
Furthermore, the generator $U(1)_V$ can be identified with the $B-L$:
\be
B-L = V= -\frac{1}{5} (Z-24X), \label{BLnumber}
\ee
which implies that $U(1)_{B-L}$ subgroup of $SO(10)$  is orthogonal
to its $SU(2)_R$ subgroup.

From the decomposition~(\ref{decomp16}) of the ${\bf 16}$ of $SO(10)$
under the maximal subgroup  $SO(10) \supset SU(5) \times U(1)_Z$ as well
as the  decomposition~(\ref{16su4}) under the 
maximal subgroup $SO(10) \supset SU(4)_{PS} \times SU(2)_L \times SU(2)_R$
we find that the $SU(5)$ multiplets ${\bf 5^*}$,${\bf 10}$ and
${\bf 1}$ have the content
\begin{eqnarray}
{\bf 5^*}_3 & = & ({\bf 3^*},{\bf 1}, I_{3R} = -1/2)_{-1/3}
                  +({\bf 1},{\bf 2},{\bf 1})_{-1},\label{5*su4}  \\
{\bf 10}_{-1} & = & ({\bf 3},{\bf 2},{\bf 1})_{1/3}
                    +({\bf 3^*},{\bf 1},I_{3R}=1/2)_{-1/3}
                    +({\bf 1},{\bf 1},I_{3R}=-1/2)_{1}, \label{10su4_1}\\
{\bf 1}_{-5} & = & ({\bf 1},{\bf 1},I_{3R} = 1/2)_1. \label{1su4}
\end{eqnarray}
under $SU(3)_C \times SU(2)_L \times SU(2)_R \times U(1)_{B-L}$.
Similarly, from the decomposition~(\ref{decomp10})  of 
the ${\bf 10}$ of $SO(10)$, we find that the multiplets
${\bf 5}$ and ${\bf 5^*}$ have the content
\begin{eqnarray}
{\bf 5}_2 & = & ({\bf 3},{\bf 1},{\bf 1})_{-2/3}
                +({\bf 1},{\bf 2},I_{3R}= -1/2)_0,  \\
{\bf 5^*}_{-2} & = & ({\bf 3^*},{\bf 1},{\bf 1})_{2/3}
                     +({\bf 1},{\bf 2},I_{3R}= 1/2)_0,
\end{eqnarray}
under $SU(3)_C \times SU(2)_L \times SU(2)_R \times U(1)_{B-L}$.
Eqs.~(\ref{5*su4}) - (\ref{1su4}) show that the embedding 
$SO(10) \supset SU(5) \supset SU(3)_C \times SU(2)_L$ is not unique.
As long as $U(1)_Y$ is not defined, there is freedom of $SU(2)_R$ rotation.
The hypercharge of the SM is not identical with $U(1)_V$ and must be orthogonal
to $SU(3)_C$ and $SU(2)_L$. This fact shows that the $U(1)_Y$ is not
orthogonal to $SU(2)_R$. Once the assignment of hypercharge is made,
the freedom of  $SU(2)_R$ rotation is eliminated. The hypercharge
assignments (\ref{gg1}) and (\ref{flipped1}) can now
be expressed in terms of the third
component of $SU(2)_R$ and the quantum number of $U(1)_V$ as
\begin{eqnarray}
Y & = & X 
    = -I_{3R} + \frac{1}{2}V,
\label{gg2}
\end{eqnarray}
for the Georgi-Glashow model and
\begin{eqnarray}
Y & = & -\frac{1}{5}(X+6Z)
               =  I_{3R} + \frac{1}{2}V,
\label{flipped2}
\end{eqnarray}
for the case of flipped emebedding. We note that hese two assignments 
differ from each other
in only the sign of the third component of $SU(2)_R$.
This means that the $SU(5)$ group of flipped model is obtained
from that of Georgi-Glashow model by the $\pi$ rotation in
$SU(2)_R$. In other words 
particle assignment of the flipped $SU(5)$ model is
obtained from that of the Georgi-Glashow $SU(5)$ model
by ``flipping'' of the $SU(2)_R$ doublets
\begin{eqnarray}
 u^c \leftrightarrow d^c, \quad e^c \leftrightarrow \nu^c.
\label{gg3}
\end{eqnarray}
Although the way of embedding $SU(5)$ as $SO(10) \supset SU(5) \times U(1)_Z
\supset SU(3)_C \times SU(2)_L \times U(1)_Y$ is not unique, the
freedom of $SU(2)_R$ rotation is no longer available now .
Thus, there are only two possibilities of embedding 
$SU(5)$ in $SO(10)$, i.e. the Georgi-Glashow $SU(5)$ or the ``flipped''
$SU(5)$.

\section{\label{section:Rgsumrules} $SO(10)$ breaking and Renormalization
Group Equations}

We now come to the question of the implications of the different patterns
of the breaking of supersymmetric $SO(10)$ to the minimal supersymemtric
standard model, based on the SM gauge group,  for the sparticle spectrum.
This can be addressed by studying the renormalization group evolution  
to the  electroweak scale.  
For the squarks and sleptons of the first and second 
family~(the light generations), the renormalization group~(RG) equations 
for the soft scalar masses are given by
\bea
16 \pi^2 {d \msQ\over dt} &=&
-{32\over 3} g_3^2 M_3^2 - 6 g_2^2 M_2^2 - {2\over 15} g_1^2 M_1^2
+ {1\over 5} g_1^2 S,
\label{rg1}
\\
16 \pi^2 {d \msuR \over dt} &=&
-{32\over 3} g_3^2 M_3^2 - {32\over 15} g_1^2 M_1^2
- {4\over 5} g_1^2 S,
\label{rg2}\\
16 \pi^2 {d \msdR \over dt} &=&
-{32\over 3} g_3^2 M_3^2 - {8\over 15} g_1^2 M_1^2
+ {2\over 5} g_1^2 S,
\label{rg3}\\
16 \pi^2 {d \mslL \over dt} &=&
- 6 g_2^2 M_2^2 - {6\over 5} g_1^2 M_1^2
- {3\over 5} g_1^2 S,
\label{rg4}\\
16 \pi^2 {d \mseR\over dt} &=&
- {24\over 5} g_1^2 M_1^2
+ {6\over 5} g_1^2 S,
\label{rg5}
\eea
where $t \equiv {\rm ln}(Q/Q_0)$, with $Q_0$ being some initial large
scale; $M_{3,2,1}$ are the running gaugino masses, $g_{3,2,1}$ are the 
usual gauge couplings associated with the SM gauge group, with 
$\alpha_i \equiv g_i^2/4\pi$,  and 
\be
S \equiv {\rm Tr}(Ym^2) = m_{H_u}^2 - m_{H_d}^2 + \sum_{\rm families}
(\msQ - 2 \msuR + \msdR - \mslL + \mseR)\> .
\ee
The $U(1)_Y$ gauge coupling $g_1$ (and $\alpha_1$)  is taken to be in
a GUT normalization throughout this paper. The quantity S evolves according to
\be
{dS \over dt} = {66\over 5} {\alpha_1\over 4 \pi}S
\ee
which has the solution
\be
S(t) = S(t_G) {\alpha_1(t) \over \alpha_1(t_G)}.
\ee
We note that if $S=0$ at the initial scale, which would be the case if
all the soft sfermion and Higgs masses are same, then the RG evolution 
will maintain it to be zero at all scales.  

The solution for the 
renormalization group equations (\ref{rg1})--(\ref{rg5})  can then  
be written as
\bea
\msuL(t) &=& \msQ(t_G) +  C_3 + C_2 + {1\over 36} C_1
+ ({1\over 2} - {2\over 3} \st) M_Z^2 \cos (2 \beta) - {1\over 5} K,
\label{sol1} \\
\msdL(t) &=& \msQ(t_G) +  C_3 + C_2  + {1\over 36}  C_1
+ (-{1\over 2} +{1\over 3} \st) M_Z^2 \cos (2 \beta)- {1\over 5} K,
\label{sol2} \\
\msuR(t) &=& \msuR(t_G) + C_3                  + {4\over 9}  C_1
+ {2\over 3} \st M_Z^2 \cos (2 \beta) + {4\over 5} K,
\label{sol3} \\
\msdR(t) & =& \msdR(t_G) +  C_3                 + {1\over 9} C_1
- {1\over 3} \st M_Z^2 \cos (2 \beta)- {2\over 5} K,
\label{sol4} \\
\mseL(t) & =& \mslL(t_G)                + C_2 + {1\over 4}   C_1
+ (-{1\over 2} + \st)  M_Z^2 \cos (2 \beta)  + {3\over 5} K,
\label{sol5} \\
\msnL(t) & = & \mslL(t_G)               +  C_2 + {1\over 4}  C_1
+ {1\over 2} M_Z^2 \cos (2 \beta) + {3\over 5} K,
\label{sol6}\\
\mseR(t) & =& \mseR(t_G)                            +   C_1
-\st M_Z^2 \cos (2 \beta) - {6\over 5} K, \label{sol7}
\eea
where $C_1,\, C_2$ and $C_3$ are given by
\bea
C_i(t)= {a_i\over 2 \pi^2} \int_t^{t_G} dt~ g_i(t)^2~ M_i(t)^2, \, i=1,2,3
\label{cidef} \\
a_1={3\over 5}, a_2={3\over 4}, a_3={4\over 3},
\eea
and
\bea
K = \frac{1}{16\pi^2}\int^{t_G}_t g_1^2(t)~S(t)~dt
  = \frac{1}{2b_1}S(t)\left[1 - \frac{\alpha_1(t_G)}{\alpha_1(t)}\right],
\eea
is the contribution of the non-universality parameter $S$ to the 
sfermion masses, and $b_1=-33/5$. 

The solutions of the RG equations for the soft scalar masses given above
involve the values of these masses at the initial scale~(GUT scale).
These initial values will be determined by the pattern of the breaking of
the grand unified  group to the SM gauge group. For the case 
of direct breaking of $SO(10)$ to the Standard Model gauge group,
these initial values are given by
\bea
\msQ(t_G) &=& \msuR(t_G) = \mseR(t_G) = m_{16}^2+g_{10}^2D, \label{cond1}\\
\mslL(t_G) &=& \msdR(t_G) = m_{16}^2-3g_{10}^2D, \label{cond2}\\
\mhU(t_G) &=& m_{10}^2-2g_{10}^2D, \label{cond3}\\
\mhD(t_G) &=& m_{10}^2+2g_{10}^2D,\label{cond4}
\eea
at the $SO(10)$ breaking scale $M_G$, where the normalization and sign 
of $D$ is arbitrary.  
Here $m_{16}$ and  $m_{10}$ are the common soft scalar masses, 
corresponding to the $\bf{16}$ and $\bf{10}$ dimensional representations, 
respectively of $SO(10)$, at the unification scale.
We note here that in the breaking of
$SO(10)$ the rank is reduced by one, and hence the $D$-term contribution
to the soft masses is expressed by a single parameter $D$.
 these intial values in the solutions~(\ref{sol1}) - (\ref{sol7}) 
of the renormalization
equations, and eliminating the quantities $m_{16}^2$, $g_{10}^2$, and 
$D$, we obtain the following two sum rules for the sfermion masses:
\bea
2 m_{\tilde{Q}}^2 - \msuR -\mseR &=& (C_3 + 2C_2 - {25\over 18}C_1),
\label{sum1}\\
m_{\tilde{Q}}^2 + \msdR - \mseR - m_{\tilde{L}}^2
&=& (2C_3 - {10\over 9} C_1)\label{sum2},
\eea
where we have used the notation
\bea
m_{\tilde{Q}}^2={1\over 2} (\msuL + \msdL),  \, \,  \, m_{\tilde{L}}^2
={1\over 2} (\mseL + \msnL). \nonumber
\eea
We note that $g_{10}^2$ and $D$ enter in the combination 
$g_{10}^2 D$ in the initial conditions~(\ref{cond1}) - (\ref{cond4}), and 
therefore constitute only one parameter. We note from the above that 
$S(t_G)=-4 g_{10}^2 D$.
The solution for $K$ is obtained by eliminating
$C_1$, $C_2$, $C_3$, $m_{16}^2$, and $m_{10}^2$  from the sfermion
mass equations. 
For the case
of direct breaking of $SO(10)$  to the SM gauge group, we have 
\bea
K &=& -{1\over 4}
(m_{\tilde{Q}}^2 - 2\msuR + \msdR +
\mseR - m_{\tilde{L}}^2 + {10\over 3}
\sin^2 \theta_W M_Z^2 \cos 2\beta)\label{su5}.
\eea
The right hand side of the sum rules (\ref{sum1}) and (\ref{sum2})
involve the functions $C_i(t)$.
These functions can be written in terms of quantities whose values can be
inferred from experiment. In terms of the gluino mass $M_{\tilde g}
= M_3(t_{\tilde g})$, we can write $C_3(t)$  from (\ref{cidef}) as
\be
C_3(t) =  {8\over 9}   {M_{\tilde g}^2\over \alpha_3^2(t_{\tilde g})} 
[\alpha_3^2(t) - \alpha_3^2(t_G)], \label{c3eqn}\\
\ee
where we have used the fact that gaugino masses run as 
\be
{M_i(t)\over \alpha_i(t)} = {M_i(t_G)\over \alpha_i(t_G)}.
\label{gauginoeq}    
\ee
In an underlying grand unified theory, we can require that all three 
gaugino masses be same at the high mass 
scale $M_G$ so that  $M_i(t_G) \equiv m_{1/2}$. We then have
\be
{M_1(t)\over \alpha_1(t)} = {M_2(t)\over \alpha_2(t)} = {M_3(t)\over
\alpha_3(t)} = {m_{1/2}\over \alpha_G},  \label{gauginogut}
\ee
where $\alpha_1(t_G) = \alpha_2(t_G) = \alpha_3(t_G) \equiv \alpha_G$
is the grand unified gauge coupling.
We note that the gaugino masses always satisfy the
relation~(\ref{gauginogut}) 
irrespective of the breaking pattern~\cite{KMYPLB} to the 
Standard Model gauge group if the underlying gauge group is
unified into a simple group at a high mass scale $M_G$. 
We further note that (\ref{gauginogut})
is a result of one-loop renormalization group equations, and does not
hold at the two loop level~\cite{yamada1}. However, the two-loop
effect is numerically small~\cite{yamada2}. 
From (\ref{gauginoeq}) it follows that
\be
M_i(t) = \alpha_i(t) {M_{\tilde g}\over \alpha_3({\tilde g})}.
\label{gauginorel}
\ee
Using the above, we can now express the functions $C_1$ and $C_2$
in terms of the gluino mass and the corresponding gauge couplings.
We have\cite{MR1}
\bea
C_1(t) & = & {2\over 11}{M_{\tilde g}^2\over \alpha_3^2({\tilde g})}
[\alpha_1^2(t_G) - \alpha_1^2(t)], \label{c1eqn}\\
C_2(t) & = & {3\over 2}{M_{\tilde g}^2\over \alpha_3^2({\tilde g})}
[\alpha_2^2(t_G) - \alpha_2^2(t)]. \label{c2eqn}
\eea
We note here that the gluino mass in (\ref{c3eqn}),\,(\ref{c1eqn})
and (\ref{c2eqn}) is the one-loop gluino mass and not the pole
mass, although these are related. 
Using these results for $C_i$, we can write the sum rules (\ref{sum1})
and (\ref{sum2}) as follows:
\bea
2 m_{\tilde{Q}}^2 - \msuR -\mseR &=& 
{M_{\tilde g}^2\over \alpha_3^2(t_{\tilde g})}
[{8\over 9} \alpha_3^2(t) - 3 \alpha_2^2(t) + {25\over 99}\alpha_1^2(t)
+ {184 \over 99} \alpha^2_G], \label{sum1new}\\
m_{\tilde{Q}}^2 + \msdR - \mseR - m_{\tilde{L}}^2
&=& 
{M_{\tilde g}^2\over \alpha_3^2(t_{\tilde g})}
[{16\over 9} \alpha_3^2(t) + {20\over 99}\alpha_1^2(t)
- {196 \over 99} \alpha^2_G].
\label{sum2new}
\eea
Using a supersymmetric threshold of
$1$~TeV, and the values $M_G = 1.9 \times 10^{16}$ GeV, $\alpha_G = 0.04$,  
$\alpha_1(1~TeV) = 0.0173, \alpha_2(1~TeV) = 0.0328, \alpha_3(1~TeV) = 0.091$,
we can finally write our sum rules in terms of 
experimentally measurable masses as~(at a scale of $1$~TeV)
\bea
2 m_{\tilde{Q}}^2 - \msuR -\mseR &=& 0.85 M_{\tilde g}^2,
\label{sum1fin}\\
m_{\tilde{Q}}^2 + \msdR - \mseR - m_{\tilde{L}}^2 & = &
1.42 M_{\tilde g}^2. \label{sum2fin}
\eea

We now come to the case of $SO(10)$ breaking via its other maximal 
subgroup $SO(10) \supset SU(4)_{PS} \times SU(2)_L \times SU(2)_R$
As in the case of breaking via the $SU(5)$ subgroup, there
appear $D$-term contributions to the soft scalar masses when the rank
of the gauge group reduces from 5 to 4 at the intermediate Pati-Salam
symmetry breaking scale $M_{PS}$. As discussed in
Section~\ref{section:so10break} matter multiplets belong either to
$L=({\bf 4}, {\bf 2}, {\bf 1})$ or $R=({\bf \bar 4}, {\bf 1}, {\bf 2})$
representations, with masses $m_L^2$ and $m_R^2$ respectively above
$M_{PS}$. When the  $SU(4)_{PS} \times SU(2)_L \times SU(2)_R$
group breaks to $G_{SM}$, we obtain the following masses
\begin{eqnarray}
\msQ(M_{PS}) &=& m_L^2 + g_4^2 D, \label{ps1}\\
\msuR(M_{PS}) &=& m_R^2 - (g_4^2 - 2 g_{2R}^2) D,  \label{ps2}\\
\mseR(M_{PS}) &=& m_R^2 + (3 g_4^2 - 2 g_{2R}^2) D,  \label{ps3}\\
\mslL(M_{PS}) &=& m_L^2 - 3 g_4^2 D, \label{ps4} \\
\msdR(M_{PS})  &=& m_R^2 - (g_4^2 + 2 g_{2R}^2) D,  \label{ps5}
\end{eqnarray}
at the Pati-Salam breaking scale. Here $D$ represents  
the $D$-term contributions whose normalization is
arbitrary. We note that these  expressions do not depend on a particular 
choice of the Higgs representation which breaks the Pati-Salam group,
and is fixed only by the symmetry breaking pattern.
We further note that the gauge coupling $g_4^2$, $g_{2R}^2$ can be determined 
from the low-energy gauge coupling $\alpha_i (m_Z)$ $(i=1,2,3)$ as
a function of $M_{PS}$ alone. The solutions~(\ref{sol1}) - (\ref{sol7})
of the renormalization
group equations~(\ref{rg1}) - (\ref{rg5}), together with the 
boundary conditions~(\ref{ps1}) - (\ref{ps5}) lead to the sum rule
\bea
m_{\tilde{Q}}^2 + \msdR - \mseR - m_{\tilde{L}}^2
&=& (2C_3 - {10\over 9} C_1)\label{sum3} \nonumber \\
&=& 
{M_{\tilde g}^2\over \alpha_3^2(t_{\tilde g})}
[{16\over 9} \alpha_3^2(t) + {20\over 99}\alpha_1^2(t)
- {196 \over 99} \alpha^2_G],
\eea
which is the only sum rule valid in this case. The notation is same as
in the case of direct breaking of $SO(10)$ to the SM gauge group.
As in the case of direct breaking of $SO(10)$, this can be
written as
\bea
m_{\tilde{Q}}^2 + \msdR - \mseR - m_{\tilde{L}}^2
=1.42 M_{\tilde g}^2.  \label{sum3fin}
\eea
Thus, this sum rule serves as a crucial
distinguishing feature of $SO(10)$ breaking via the Pati-Salam subgroup.
If both the sum rules~(\ref{sum1fin}) and ~(\ref{sum2fin}) are seen to hold
experimentally, then in the context of $SO(10)$ unification, the breaking
of $SO(10)$ takes place directly to the SM gauge group. 
On the other hand, if only
the sum rule~(\ref{sum3fin}) is seen to hold experimentally, then the breaking
of $SO(10)$ must take place  via the Pati-Salam subgroup.
We recall the relation~(\ref{gauginogut}) that has been used in the
derivation of (\ref{sum3fin})
is valid irrespective of the breaking pattern~\cite{KMYPLB} to the
Standard Model gauge group if the underlying gauge group is
unified into a simple group at a high mass scale $M_G$.
Furthermore, we note that the parameter $K$ cannot be determined in the case
of breaking via the Pati-Salam subgroup.

It is important to distinguish between the situation where there
is a grand unification of the SM gauge group into a simple group 
like $SO((10)$, 
and the situation where there is no such unification into a simple group. 
A typical example
of the latter case is the flipped $SU(5)\times U(1)$ model which is
not grand-unified into a simple  group. In this case we have two independent
gauge couplings $g_{SU(5)}$ and $g_{U(1)}$ at the GUT scale. On the other 
hand there are three soft scalar masses ${{m}_{10}}^2$, ${{m}_{\bar{5}}}^2$, 
and ${{m}_{1}}^2$ at the GUT scale. In addition there is the unknown $D$-term.
The intial values of the soft scalar masses are given by
\bea
\msQ(t_G) &=& {{m}_{10}}^2 + \left({1 \over 10}{g_{SU(5)}}^2 
+ {1 \over 40}{g_{U(1)}}^2\right) D, \label{ununi1}\\
\msuR(t_G) &=& {{m}_{\bar{5}}}^2 +
\left({1 \over 5}{g_{SU(5)}}^2 - {3 \over 40}{g_{U(1)}}^2 \right) D ,  
\label{ununi2}\\
\mseR(t_G) &=& {{m}_{1}}^2 + {1 \over 8}{g_{U(1)}}^2 D,  \label{ununi3}\\
\mslL(t_G) &=& {{m}_{\bar{5}}}^2 - \left({3 \over 10}{g_{SU(5)}}^2 
+ {3 \over 40}{g_{U(1)}}^2 \right) D, \label{ununi4} \\
\msdR(t_G)  &=& {{m}_{10}}^2 - \left({2 \over 5}{g_{SU(5)}}^2 
- {1 \over 40}{g_{U(1)}}^2 \right) D.  \label{ununi5}
\eea
In this case, eliminating the unknown soft mass parameters
${{m}_{10}}^2$, ${{m}_{\bar{5}}}^2$,
and ${{m}_{1}}^2$,  the  gauge couplings
$g_{SU(5)}$ and $g_{U(1)}$, and the parameter  $D$ 
from the solutions~(\ref{sol1}) - (\ref{sol7})
of the renormalization
group equations~(\ref{rg1}) - (\ref{rg5}), together with the
boundary conditions~(\ref{ununi1}) - (\ref{ununi5}), we get the 
sum rule 
\bea
m_{\tilde{Q}}^2 - \msuR - \msdR + m_{\tilde{L}}^2
&=& -(C_3 - 2 C_2 + {5\over 18} C_Y), \label{sum4}
\eea
where, to avoid confusion, we have denoted the function $C$
corresponding to the $U(1)_Y$ subgroup of the Standard Model
as $C_Y$~(with the usual GUT normalization). 
That this can be done is a consequence of the general argument for
elimination of $D$-terms.

We can now try to write the right hand side of (\ref{sum4}) in terms of
measurable quantities, just as we did in the case of the unified gauge
group $SO(10)$. To do this, we note that the value of $g_{SU(5)}$ 
at the scale $M_G$ is the same as in the case of 
a grand unified supersymmetric gauge theory like
$SO(10)$, i.e.
$g^2_{SU(5)}(M_G)/(4\pi) = \alpha_G$, since it is determined by
the evolution of $SU(3)$ and
$SU(2)$ gauge couplings, which is unaltered.
This implies
\be
{M_2(t)\over \alpha_2(t)} = {M_3(t)\over
\alpha_3(t)} = {m_{1/2}\over \alpha_G}.  \label{gauginonongut}
\ee
Furthermore, we can again identify the gluino mass as 
$M_{\tilde g} = M_3(t_{\tilde g})$. We then have
\bea
C_3(t) & = &  {8\over 9}   {M_{\tilde g}^2\over \alpha_3^2(t_{\tilde g})} 
[\alpha_3^2(t) - \alpha_3^2(t_G)], \label{c3eqnng}\\
C_2(t) & = & {3\over 2}{M_{\tilde g}^2\over \alpha_3^2({\tilde g})}
[\alpha_2^2(t_G) - \alpha_2^2(t)]. \label{c2eqnng}
\eea
On the other hand, we can write the function $C_Y$ as 
\bea
C_Y(t) & = & {2\over 11}{M_Y^2(t)\over \alpha_Y^2(t)}
[\alpha_Y^2(t_G) - \alpha_Y^2(t)],  \label{cYeqnng}
\eea
where we have denoted the soft gaugino mass corresponding to the $U(1)_Y$
gauge group as $M_Y$. Using (\ref{c3eqnng}),  (\ref{c2eqnng}) and
(\ref{cYeqnng}) we can write  the sum rule (\ref{sum4}) as
\bea
m_{\tilde{Q}}^2 - \msuR - \msdR + m_{\tilde{L}}^2
& = & 
-{M_{\tilde g}^2\over \alpha_3^2(t_{\tilde g})}
[{8\over 9} \alpha_3^2(t) + 3\alpha_2^2(t)
- {35 \over  9} \alpha^2_G]
- {5\over 99} {M_Y^2(t) \over \alpha_Y^2(t)}
[\alpha_Y^2(t_G) - \alpha_Y^2(t)].
\label{sum4new}
\eea
We note that the gaugino mass parameter $M_Y$ can be extracted from the
experimental measurements in the neutalino sector~\cite{Choi}, 
so that the sum rule (\ref{sum4new}) can be tested.
Thus, this sum rule could
serve to distinguish the flipped ununified  $SU(5)\times U(1)$ model
from the unified $SO(10)$ model.

To sum up this Section, we have shown that there are characteristic sum rules
obeyed by the sfermion masses when $SO(10)$ breaks to the SM gauge group
via different breaking chains. We have also shown that in the case the SM does 
not unify into a simple gauge group, there is a sum rule which can help 
distinguish such a situation from the one where the SM unfies into
a simple group at the grand unified scale.


\section{\label{section:Numerical} Numerical results}
In this Section we consider the phenomenological 
implications of the $D$-term contributions
that arise in the breaking of $SO(10)$ to the SM gauge group,
as well as typical non-universality associated with
the Pati-Salam subgroup for purposes of illustration.
Recalling here that one of the most attractive 
pictures of unification is the one where there is gauge coupling unification 
at a single scale, and in which Yukawa couplings of the heaviest generation 
have a common value.  This also fixes the value of the hitherto 
unknown parameter $\tan\beta \equiv <H_u^0/H_d^0>$~(the ratio of the vacuum 
expectation values of the two Higgs doublets of the minimal supersymmetric
standard model~(MSSM)) at the theoretically attractive value 
of $\sim m_t/m_b$.  
Furthermore, this framework  also provides a candidate for
the cold dark matter~\cite{BHS} of the universe in the form of a bino-like
lightest supersymmetric particle~\cite{AS1}.
However, this simple picture is not realised when threshold 
corrections,  that depend crucially on the details of the spectrum,  
are taken into account~\cite{BDR}.  Furthermore, it is  well known
that the model has problems of naturalness when the tree
level Higgs potential is considered, although arguments have been presented 
to show that one-loop corrections might alleviate the problem~\cite{BFT}.
In other words, one may have to give up exact unification, but 
still have large values of $\tan\beta$ and approximate unification, 
or alternatively constrain the parameter
space significantly by demanding exact unification.  Our approach here
will be to take the simplest possible set of assumptions and study
the implications of these for the phenomenology.  This could, then,  
become a basis for further studies, and could  become 
important when the  supersymmetric particles are discovered
experimentally.

In order to implement the above picture, we carry  out 
numerical integration of the renormalization group equations 
for  the gauge couplings, Yukawa couplings, the gaugino masses, the
supersymmetry breaking soft scalar mass squared parameters and the 
soft trilinear couplings of
the MSSM with $SO(10)$ breaking boundary conditions.  
For definiteness, we shall consider the case of $SO(10)$ breaking
via the Pati-Salam subgroup, since this case has more parameter freedom.
We wish to retain those aspects of unification that
are approximately valid.  Motivated by $SO(10)$ unification,
these include  a unified gauge coupling  at $M_X$,
and a unified Yukawa coupling for the heaviest generation.  
For all parameters except the mass squared
parameters, which we study case by case,
we assume universal boundary conditions.
Starting with values of the common gaugino mass ($M_{1/2}$),
trilinear couplings ($A$), with the third generation
Yukawa couplings having a common value $h$ at the unification scale, 
and the unified gauge coupling$\alpha_G$,  we  integrate
the set of coupled renormalization group equations from the $SO(10)$ 
breaking scale down to the effective supersymmetry breaking scale 
of $\sim 1$ TeV.  
At this scale, the parameters in the Higgs potential, after being evolved 
from the GUT breaking scale, must be such that the electroweak symmetry
is broken. As is well known, one of the conditions for this to happen is
\begin{equation}
{\mu_1^2 -  \mu_2^2  \tan^2 \beta \over \tan^2\beta -1}= {m_Z^2\over 2},
\end{equation}
where $\mu_1^2=m_{H_d}^2 + \mu^2$ and $\mu_2^2=m_{H_u}^2 + \mu^2$,
with $m_{H_u}^2$ and $m_{H_d}^2$ being  soft supersymmetry breaking
Higgs mass squared parameters, and 
$\mu$  the supersymmetry conserving Higgs(ino) mass parameter. 
Proceeding in the by now well-known fashion~\cite{AS1} of determining 
$\tan\beta$ from the accurately known value of the $\tau$-lepton mass,  
inserting it into the above equation, and using the values of
the Higgs mass squared parameters determined from the RG evolution, 
yields the parameter $\mu$.
We chose the  sign of $D$ to be positive.  This
alleviates the problem of fine-tuning, inherent in  $SO(10)$ unification, by
allowing $m_{H_u}^2$ to evolve to values that are negative and
larger in magnitude, compared to the situation when the  $D$-term 
is absent.

We note  that for the universal mass squared case,
sufficiently large values
of the common gaugino mass $M_{1/2}$ are required to ensure that 
the gluino is sufficiently heavy, 
and also fairly large values of the common soft scalar mass
$m_0$ ($< M_{1/2}$)
are required to ensure that the neutralino~(and not the lightest slepton)
is the LSP.  The near degeneracy of the two slepton states in the
absence of electroweak symmetry breaking makes the mixing between them,
once $SU(2)\times U(1)$ is broken, significant.  
Indeed, it is important to observe the variation of the mass of the 
lighest slepton, since it has
the tendency to become lighter than the lightest neutralino and to emerge
as a candidate for the LSP, which is not acceptable.  
An upper bound on $m_0$ ensues when we  require  
a sufficiently large $m_A \equiv \mu_1^2+\mu_2^2 (\gsim M_Z)$.
Keeping these features in mind, we study the effects of the $D-$term
on the spectrum.

\begin{figure}[hbt]
\begin{center}
\includegraphics[width=10cm]{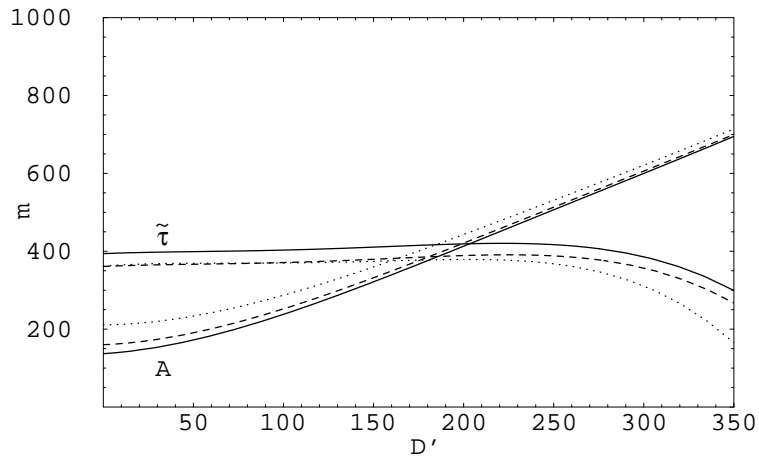}
\caption{\label{fig:1} 
Values of $m_A$ and lighter stau~($\tilde{\tau}$) mass  plotted as
a function of $D'$ with $M_{1/2}=800, A=0$ and $\delta m_0^2=-(200)^2$ for
the case of $SO(10)$ breaking via Pati-Salam subgroup.
The solid line corresponds to $m_0=700, h_t=h_b=h_\tau=2$,
dashed line corresponds to $m_0=700, h_t=h_b=h_\tau=3$,
and dotted line to $m_0=600, h_t=h_b=h_\tau=2$. All masses are in GeV. }
\vspace{-.7cm}
\end{center}
\end{figure}

\bigskip

\begin{figure}[hbt]
\begin{center}
\includegraphics[width=10cm]{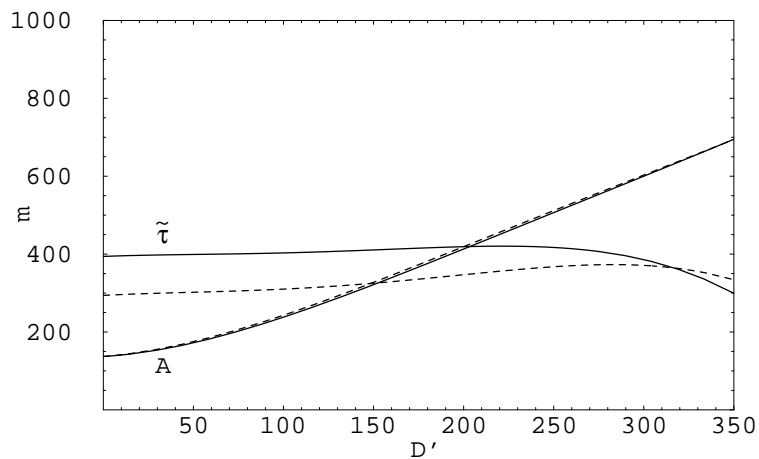}
\caption{\label{fig:2} 
Values of $m_A$ and lighter stau~($\tilde{\tau}$) mass  plotted as
a function of $D'$ for $M_{1/2}=800, A=0$ and $m_0=700, h_t=h_b=h_\tau=2$ for
the case of $SO(10)$ breaking via Pati-Salam subgroup.
Here the solid line corresponds to
$\delta m_0^2=-(200)^2$ and dashed line corresponds to 
$\delta m_0^2=200^2$.}
\vspace{-.7cm}
\end{center}
\end{figure}

We start by determining the low-energy sparticle spectrum characteristics
for a set of typical boundary conditions appropriate to the breaking of
$SO(10)$ via the Pati-Salam subgroup.
For this case, we have imposed the
condition $g_{4}=g_{2 R}=g_{2 L}(=g_{10})$ at
the GUT scale.  We consider the following cases:
$m_h^2=m_0^2,\, m_R^2=m_0^2-\delta m_0^2$, $m_L^2=m_0^2+ \delta m_0^2$.  
With these boundary conditions, we expect significant changes in the
masses of the squarks compared to the situation when there are universal
boundary conditions for the squark masses. What is relevant here, however,
is the influence on the masses of the stau's, because of the
mixing between the left and right states after electroweak
symmetry breaking.  Normally part of the allowed parameter space
is ruled out since the lighter stau tends to become lighter
than the lightest neutralino, which is unacceptable on phenomenological
grounds. We illustrate these features through our numerical results.  

In Fig.~\ref{fig:1}, we illustrate the variation of the low energy
observables as the parameter $D'\equiv \sqrt{g_{10}^2 D}$ is
varied in the range of 0 to 350 GeV, where $\delta m_0^2$ is taken 
to be $-(200 {\rm GeV})^2$, for
a typical parameter choice of $M_{1/2}=800,\, m_0=700,\, A_0=0$ in units
of GeV, with the common Yukawa coupling taking the value
$h_t=h_b=h_\tau=2.0$. This is shown as a solid line in Fig.~\ref{fig:1}
We have also considered the case when all parameters 
take the above quoted values, except that the unified Yukawa coupling is taken
as 3, and also for the case when only $m_0$ is changed to a value of 600 GeV.
We recall that one of the stringent constraints 
is the requirement that the mass of $\tilde{\tau}$ exceed the lightest
neutralino mass, which in the present case is almost entirely bino-like
with a mass $\sim 350$ GeV.  This implies that
the larger value of $m_0$ is preferred.  Thus a window of parameters
is allowed when D-term non-universality contributes significantly
to the boundary conditions.  Note that we have not imposed any
constraints on bino-purity which typically constrains larger values
of the unified Yukawa coupling, since the possibility of Higgsino
like dark matter is not excluded~\cite{CCN}.

In Fig.~\ref{fig:2} we have illustrate the numerical results
for a typical
choice of parameters with  $\delta m_0^2=\pm (200 {\rm GeV})^2$.
It may be inferred from here that the negative sign is preferred for
lower values of D-term non-universality before the onset of strong mixing
between the scalar leptons of the heaviest generation, while the converse
is true for higher values of D-term non-universality. We conclude
that with the non-universality coming from the $D$-term contribution
via $SO(10)$ breaking, the naturalness problems are alleviated by the 
presence of additional parameters. The main reason for this
is that such a $D-$ term non-universality succeeds in splitting the masses 
of the Higgs doublets at the electroweak scale in an efficient manner.

\section{Summary and conclusions}

In this paper we have considered the breaking of $SO(10)$ grand unfied gauge 
group to the SM gauge group in a supersymmetric grand unified theory.
Such a  breaking of $SO(10)$ generates non-universal contributions
to the soft scalar masses. We have studied the implications
of such non-universal contributions for the sfermion masses. In particular,
we have derived sum rules which hold among the sfermion masses
when  $SO(10)$ breaks to the SM model via different breaking chains. 
These sum rules may help in distinguishing between the different
breaking patterns. We have also shown that these sum rules are different
from the case when the SM is not unified into a simple group.
We have studied the implications of the non-universal contributions
to the low energy phenomenology. In particular, we have shown that 
it is possible to increase  the unified Yukawa coupling
to as large a value as 3, once D-term non-universality as well
as additional non-universality is introduced.  
Typically,  in the absence of such a non-universality
it is not possible to achieve Yukawa unification and radiative electroweak
symmetry breaking, and to satisfy all the phenomenological constraints
for  typical values of parameters.
With the non-universality coming from the $D$-term contributions
and additional non-universality, the naturalness problems associated 
with large values of  $\tan\beta$ are alleviated by the
presence of additional parameters. 

\newpage

{\bf Acknowledgements:} BA thanks the Department of Science and
Technology,  and Council for Scientific and Industrial Research,
India, whereas
PNP thanks the University Grants Commission, and Council for
Scientific and Industrial Research, India,
for support during the course of this work.


\end{document}